\documentclass[11pt,a4paper]{article}

\usepackage[margin=2.5cm]{geometry}
\usepackage{amsmath,amssymb,amsthm,mathtools,bm}
\usepackage{hyperref}
\usepackage{graphicx}
\usepackage{booktabs}

\title{Admissible Reconstruction of Reaction-Channel Levels on Fixed Subgroup Support for Cross-Section-Space Probability Table Constructions}

\author{
Beichen Zheng\thanks{Corresponding author}, Lili Wen\\[0.5em]
\small China Nuclear Data Center, China Institute of Atomic Energy, Beijing, China
}

\date{}

\begin{document}

\maketitle

\begin{abstract}
In cross-section-space probability table constructions, reaction-channel levels are reconstructed on fixed total-subgroup nodes and probabilities. Although the standard full-matching reconstruction is uniquely determined, it does not in general preserve componentwise nonnegativity of the channel levels. We impose nonnegativity both for physical interpretability and because, on fixed positive total-subgroup nodes and probabilities, it provides a sufficient structural condition for nonnegativity of the folded effective
cross section over all dilutions. We therefore formulate an admissible constrained reconstruction problem on the fixed subgroup support, in which selected low-order channel information is retained exactly and the remaining matching conditions are fitted in a weighted least-squares sense. After null-space reduction, the problem becomes a convex optimization problem with linear inequality constraints. For the single-retention formulation, nonnegative feasibility is automatic when the retained \(0\)-order aggregate is nonnegative, whereas for a two-retention variant it additionally requires a compatibility condition with the fixed total-subgroup nodes. Numerical results for a representative \(^{238}\mathrm{U}\) capture benchmark show that nonnegativity violations are confined to a small subset of energy groups. On these groups, the admissible reconstruction restores nonnegativity, but at the cost of some response-level deterioration relative to full matching. In the comparison, the single-retention formulation shows the more stable overall behavior.
\end{abstract}


\section{Introduction}

Resonance self-shielding is one of the main sources of strong within-group
cross-section heterogeneity in multigroup reactor calculations. To represent
this behavior at acceptable computational cost, the subgroup
method replaces the underlying continuous variation by a finite set of
representative levels together with associated weights or probabilities
\cite{Levitt1972ProbabilityTable,HebertCoste2002MomentProbabilityTables,Hebert2005Ribon,ChibaUnesaki2006MomentProbability}.
Over time, this general line of methodology has been extended beyond its
classical forms in several directions. On the one hand, the framework has been
recast for broader transport settings and further developed through subgroup
projection formulations \cite{Hebert1997SubgroupArbitraryGeometry,Hebert2009SubgroupProjection}.
On the other hand, continued work has refined the construction and use of the
reduced subgroup representation itself, including improved subgroup-weight
generation and finer subgroup formulations such as fine-group and fine-mesh
variants \cite{JooKimPogosbekyan2009SubgroupWeights,LiZhangZhangZhao2020FineGroupSlowingDown,LiEtAl2023FineMeshSubgroup}.
Related developments have also introduced temperature-dependent formulations on
the same general subgroup basis \cite{JungLimJoo2016TemperatureDependentSubgroup}.
Across these formulations, the construction ultimately delivers a reduced subgroup representation for subsequent use. Besides reproducing the target effective cross sections accurately, this representation must also remain admissible in the downstream subgroup and transport calculations.

In an \(N\)-level representation, the subgroup data comprise total subgroup
levels, subgroup probabilities, and reaction-channel levels. The total
subgroup levels together with the subgroup probabilities define the reduced
subgroup representation, while the reaction-channel levels provide
channelwise detail on that representation. The most basic admissibility
requirements in this setting concern three quantities: the nonnegative character
of the subgroup quantity entering the subgroup fixed-source transport
calculation, the nonnegativity of the subgroup probabilities as discrete
weights, and the nonnegativity of the effective cross sections ultimately
supplied to the main multigroup transport calculation. These requirements, however, are not all guaranteed at the same structural level.
Across resonance self-shielding methods, different constructions control
different parts of the reduced representation and therefore provide different
types of structural guarantees \cite{LiuMartin2017PinResolvedSelfShielding,Hebert2024RSE}.

Within the cross-section-space branch of probability table constructions, encompassing both the
moment-based route and discrete-measure-based route
\cite{RibonMaillard1986ProbabilityTables,Zheng2026GQC}, the overall workflow
follows the same broad pattern. The total-channel information is first reduced
to a finite subgroup representation, namely the total subgroup levels and
subgroup probabilities that fix the subgroup support. At this stage, the
construction remains within the Gauss-type quadrature structure underlying the
total-level compression, and the total subgroup representation inherits the
corresponding nonnegative-real structure \cite{AllenChuiMadychNarcowichSmith1974PadeGaussianQuadrature,Gautschi1968ConstructionGaussChristoffel,Gautschi1970GaussianQuadratureModifiedMoments,GolubWelsch1969GaussQuadrature}. The reaction-channel subgroup levels
are then reconstructed in a second stage. In both routes, they are commonly
recovered by exact satisfaction of the prescribed algebraic conditions on the fixed support. This choice is attractive from the standpoints of exactness and computational simplicity: it preserves the prescribed information exactly, yields a uniquely determined reconstruction, and avoids an additional constrained solve.

The difficulty is that exact full matching does not, in general, preserve
componentwise nonnegativity of the reconstructed reaction-channel levels. The
resulting fixed-node reconstruction may therefore remain algebraically exact
while still containing negative channel values, so that the folded effective
cross section is no longer
structurally guaranteed to remain nonnegative. We therefore take
componentwise nonnegativity of the reaction-channel levels as the admissibility
requirement in the present work, both for physical interpretability and because,
on fixed positive subgroup nodes and probabilities, it provides a sufficient
structural condition for nonnegativity of that folded effective cross section
for every dilution.
Earlier response-fitted probability table constructions have also imposed nonnegativity on reaction-specific subgroup cross sections; for example, Peng et al.\ impose such constraints on subgroup absorption, scattering, and fission cross sections \cite{PengJiangZhangWang2013RIF}.

In the present work, we reconstruct the reaction-channel levels directly on a fixed total subgroup support, taking the full-matching solution as the reference. When that solution is already admissible, it is retained unchanged; otherwise, it is replaced by an admissible alternative that preserves selected low-order channel information exactly while fitting the remaining conditions under a nonnegativity requirement. This yields an admissible reconstruction for fixed-node reaction-channel levels, especially natural for discrete-measure-based probability table constructions \cite{Zheng2026GQC}, in which the fixed total-subgroup rule is generated through a Lanczos--Jacobi route \cite{MeurantStrakos2006LanczosFinitePrecision,GolubWelsch1969GaussQuadrature} and is already available in a numerically nonnegative-real form.

The remainder of the paper is organized as follows. Section~2 formulates the
fixed-node reaction-channel reconstruction problem and analyzes the
nonnegativity failure of exact full matching. Section~3 presents the admissible
reconstruction under exact retention of selected low-order information.
Section~4 reports numerical results, and Section~5 concludes the paper.

\section{Problem Setting and Nonnegativity Failure of Full Matching}

In probability table constructions, a resonance group is represented
by total subgroup levels, subgroup probabilities, and reaction-channel subgroup
levels. We consider the second stage of this workflow: the reconstruction of
reaction-channel subgroup data on an already fixed total-subgroup rule.

For a given energy group \(g\) and a fixed reaction channel \(x\), the target
quantity is the effective cross section
\begin{equation}
\sigma^{\mathrm{ref}}_{x,g}(\sigma_0)
=
\frac{\displaystyle \int_{E_{g+1}}^{E_g} \sigma_x(E)\,\phi(E;\sigma_0)\,dE}
{\displaystyle \int_{E_{g+1}}^{E_g} \phi(E;\sigma_0)\,dE},
\label{eq:spr_exact_effective_xs}
\end{equation}
where \(\phi(E;\sigma_0)\) denotes the within-group flux under dilution
\(\sigma_0>0\). Under the narrow-resonance approximation, up to an
energy-independent normalization,
\[
\phi(E;\sigma_0)\propto \frac{1}{\sigma_t(E)+\sigma_0},
\]
so that \eqref{eq:spr_exact_effective_xs} becomes
\begin{equation}
\sigma^{\mathrm{ref}}_{x,g}(\sigma_0)
=
\frac{\displaystyle \int_{E_{g+1}}^{E_g}
\sigma_x(E)\,(\sigma_t(E)+\sigma_0)^{-1}\,dE}
{\displaystyle \int_{E_{g+1}}^{E_g}
(\sigma_t(E)+\sigma_0)^{-1}\,dE}.
\label{eq:spr_nra_reference_response}
\end{equation}

Let
\[
\{(\sigma_{t,i},p_i)\}_{i=1}^{N},
\qquad
p_i>0,
\qquad
\sum_{i=1}^{N} p_i = 1,
\]
denote the total subgroup levels and subgroup probabilities obtained from the
preceding total-subgroup compression step. In discrete-measure-based construction\cite{Zheng2026GQC}, these quantities are obtained
from the eigendecomposition
\[
J_N = Q \Lambda Q^{\mathsf T},
\qquad
\Lambda = \mathrm{diag}(\sigma_{t,1},\dots,\sigma_{t,N}),
\qquad
p_i = Q_{0i}^2 .
\]
At the present stage, \(\{(\sigma_{t,i},p_i)\}_{i=1}^N\) are fixed, and the
unknown is the vector of reaction-channel subgroup levels
\[
s := (\sigma_{x,1},\dots,\sigma_{x,N})^{\mathsf T}\in\mathbb R^{N}.
\]

Once \(s\) is specified, the corresponding subgroup prediction for the effective
cross section of channel \(x\) at dilution \(\sigma_0>0\) is
\begin{equation}
\sigma_x^{\mathrm{PT}}(\sigma_0)
=
\frac{\displaystyle \sum_{i=1}^{N} p_i \,\sigma_{x,i}\,(\sigma_{t,i}+\sigma_0)^{-1}}
{\displaystyle \sum_{i=1}^{N} p_i \,(\sigma_{t,i}+\sigma_0)^{-1}}.
\label{eq:spr_subgroup_response}
\end{equation}
For convenience, define
\begin{equation}
N_x^{\mathrm{PT}}(\sigma_0;s)
:=
\sum_{i=1}^{N} p_i\,\sigma_{x,i}\,(\sigma_{t,i}+\sigma_0)^{-1},
\qquad
D^{\mathrm{PT}}(\sigma_0)
:=
\sum_{i=1}^{N} p_i\,(\sigma_{t,i}+\sigma_0)^{-1},
\label{eq:spr_num_den}
\end{equation}
we may write
\begin{equation}
\sigma_x^{\mathrm{PT}}(\sigma_0)
=
\frac{N_x^{\mathrm{PT}}(\sigma_0;s)}{D^{\mathrm{PT}}(\sigma_0)}.
\label{eq:spr_response_ratio}
\end{equation}
Thus the total-subgroup rule is fixed, whereas the channelwise nodal values
\(s_i\) entering the numerator of \eqref{eq:spr_response_ratio} remain to be
reconstructed.

In discrete-measure-based construction, the reaction-channel levels are recovered on this fixed rule by
matching the first \(N\) orthogonal-basis coefficients. Let
\[
b_k,\qquad k=0,\dots,N-1,
\]
denote the prescribed reaction-channel coefficients. The matching conditions are
\begin{equation}
b_k = \sum_{i=1}^{N} (Q_{ki}Q_{0i})\,\sigma_{x,i},
\qquad k=0,\dots,N-1,
\label{eq:full_matching_component}
\end{equation}
or, in matrix form,
\begin{equation}
Ms=b,
\label{eq:full_matching_matrix}
\end{equation}
with
\[
M_{ki}=Q_{ki}Q_{0i},
\qquad
s=(\sigma_{x,1},\dots,\sigma_{x,N})^{\mathsf T},
\qquad
b=(b_0,\dots,b_{N-1})^{\mathsf T}.
\]

Since
\[
M = Q\,\mathrm{diag}(Q_{01},\dots,Q_{0N}),
\]
and \(p_i=Q_{0i}^2>0\) implies \(Q_{0i}\neq 0\) for all \(i\), the matrix
\(M\) is nonsingular. Therefore \eqref{eq:full_matching_matrix} admits the
unique solution
\begin{equation}
s^{\mathrm{full}} = M^{-1} b.
\label{eq:unique_full_solution}
\end{equation}

The issue is not solvability or uniqueness, but admissibility. The full-matching
solution \eqref{eq:unique_full_solution} satisfies the prescribed coefficient
conditions exactly, but it does not in general guarantee
\begin{equation}
s_i \ge 0,
\qquad i=1,\dots,N.
\label{eq:nonnegativity_constraint}
\end{equation}
That is, exact coefficient matching and componentwise nonnegativity are distinct
requirements, and the former does not imply the latter.

This distinction appears specifically at the reaction-channel reconstruction
stage. The preceding total-subgroup compression is tied to a positive-measure
construction, so the total subgroup levels and probabilities inherit a
nonnegative-real structure. By contrast, once these total quantities are fixed,
the reconstruction of reaction-channel levels on that support is no longer
automatically positivity preserving. Consequently, \(s^{\mathrm{full}}\) may be
algebraically exact and yet contain negative components.

In this work, admissibility is taken to mean componentwise nonnegativity of the
reaction-channel levels. This condition is stronger than requiring only
a nonnegative folded response, but it is sufficient for the latter. Indeed, for
every \(\sigma_0>0\),
\begin{equation}
D^{\mathrm{PT}}(\sigma_0) > 0,
\label{eq:spr_positive_denominator}
\end{equation}
and if
\begin{equation}
s_i\ge 0,\qquad i=1,\dots,N,
\label{eq:spr_nonnegative_nodes_again}
\end{equation}
then
\begin{equation}
N_x^{\mathrm{PT}}(\sigma_0;s)\ge 0
\quad\Longrightarrow\quad
\sigma_x^{\mathrm{PT}}(\sigma_0)\ge 0.
\label{eq:spr_nonnegative_nodes_imply_response}
\end{equation}

We therefore consider the admissible set
\begin{equation}
\mathcal F_{+}
:=
\left\{
s\in\mathbb R^{N} :
Es=r,\;
s_i\ge 0,\ i=1,\dots,N
\right\},
\label{eq:nonnegative_feasible_set}
\end{equation}
where \(Es=r\) represents the selected low-order channel information to be kept
exact in the admissible reconstruction. Here
\[
E\in\mathbb R^{m\times N},
\qquad
r\in\mathbb R^m,
\qquad
m\ll N.
\]
Thus the retained conditions define a low-dimensional affine subspace, while
nonnegativity imposes the admissibility constraint.

The reconstruction problem is called feasible if
\(\mathcal F_{+}\neq\varnothing\). We first compute the full-matching solution
\(s^{\mathrm{full}}\). If \(s^{\mathrm{full}}\in\mathcal F_{+}\), it is already
admissible and is accepted. Otherwise, we seek a reconstruction directly in
\(\mathcal F_{+}\). The next section specifies the retained operator \(E\) and
formulates this admissible reconstruction explicitly.

\section{Admissible Reconstruction with Exact Retention}

We formulate the channel-level reconstruction on the fixed subgroup nodes and
probabilities as an admissible constrained problem. The reconstruction seeks a
nonnegative channel representation that satisfies the selected retained
conditions exactly and fits the remaining matching conditions in a weighted
least-squares sense. The resulting problem is a constrained least-squares
problem on the affine subspace defined by the exact-retention constraints.

Let \(\{E_j\}_{j=0}^{J}\) denote the fine-state points inherited from the
union-grid realization in discrete-measure-based construction, where \(J\) is the last union-grid index, and write
\[
r_{t,j}:=\sigma_t(E_j),
\qquad
r_{x,j}:=\sigma_x(E_j),
\]
with associated nonnegative source-data weights \(\omega_j\).

\paragraph{Single-retention formulation.}
The retained quantity in the default formulation is chosen as the fine-state
channel aggregate corresponding to Chiba's \(0\)-order condition:
\begin{equation}
	m_{0}^{\mathrm{fine}}
	:=
	\sum_{j=0}^{J}\omega_j r_{x,j}.
	\label{eq:dar_m0fine}
\end{equation}
Under Chiba's prescription, this \(0\)-order quantity is associated with the
infinite-dilution endpoint. The default reconstruction therefore enforces this
quantity exactly \cite{ChibaUnesaki2006MomentProbability}.

After a possible row permutation, write the full matching system as
\begin{equation}
	Ms=b,
	\qquad
	M=
	\begin{pmatrix}
		p_1 & \cdots & p_N \\
		& C &
	\end{pmatrix},
	\qquad
	b=
	\begin{pmatrix}
		m_{0}^{\mathrm{fine}} \\
		c
	\end{pmatrix},
	\label{eq:dar_partitioned_system}
\end{equation}
where the first row represents the retained equality condition and
\(C\in\mathbb{R}^{(N-1)\times N}\), \(c\in\mathbb{R}^{N-1}\) collect the
remaining \(N-1\) matching equations.
Introduce the retained linear operator
\begin{equation}
	E_0 :=
	\begin{pmatrix}
		p_1 & \cdots & p_N
	\end{pmatrix},
	\qquad
	r_0 := m_{0}^{\mathrm{fine}}.
	\label{eq:dar_E0_r0}
\end{equation}
The exact-retention condition is therefore
\begin{equation}
	E_0 s = r_0,
	\qquad\text{i.e.}\qquad
	\sum_{i=1}^{N} p_i s_i = m_{0}^{\mathrm{fine}}.
	\label{eq:dar_exact_retention_single}
\end{equation}

For the single-retention formulation, we take the explicit particular vector
\begin{equation}
	s_p^{(0)} := m_{0}^{\mathrm{fine}}\mathbf{1},
	\qquad
	\mathbf{1}:=(1,\dots,1)^{\top}.
	\label{eq:dar_particular_single}
\end{equation}
Since \(p_i\ge 0\) and \(\sum_{i=1}^{N}p_i=1\), this choice satisfies
\begin{equation}
	E_0 s_p^{(0)} = r_0.
	\label{eq:dar_particular_single_exact}
\end{equation}
Let \(Z_0\in\mathbb{R}^{N\times(N-1)}\) have columns forming a basis of
\(\ker(E_0)\), so that
\begin{equation}
	E_0 Z_0 = 0.
	\label{eq:dar_nullspace_single}
\end{equation}
Then every vector satisfying the exact-retention condition
\eqref{eq:dar_exact_retention_single} can be written uniquely in the form
\begin{equation}
	s = s_p^{(0)} + Z_0 y,
	\qquad
	y\in\mathbb{R}^{N-1}.
	\label{eq:dar_affine_param_single}
\end{equation}
This null-space parametrization embeds the equality constraint exactly and is
standard in equality-constrained least-squares formulations
\cite{Bjorck1996LeastSquares,HaskellHanson1981EqualityNonnegativity}.
Thus the reconstruction is restricted from the outset to the affine subspace
\(\{s\in\mathbb{R}^N:E_0 s=r_0\}\), and the optimization is carried out only
over the reduced variable \(y\).

To balance the remaining matching equations, introduce positive normalization
factors \(\alpha_k>0\) and define
\begin{equation}
	\widehat C_{k,:}:=\alpha_k^{-1}C_{k,:},
	\qquad
	\widehat c_k:=\alpha_k^{-1}c_k,
	\qquad
	k=1,\dots,N-1.
	\label{eq:dar_scaled_rows_single}
\end{equation}
Let
\[
W_s=\mathrm{diag}(\eta_1,\dots,\eta_{N-1}),
\qquad
\eta_k>0.
\]
Substituting \eqref{eq:dar_affine_param_single} into the residual of the
remaining equations yields the reduced problem
\begin{equation}
	\min_{y\in\mathbb{R}^{N-1}}
	\frac12
	\bigl\|
	W_s^{1/2}\bigl(\widehat C(s_p^{(0)}+Z_0 y)-\widehat c\bigr)
	\bigr\|_2^2
	\label{eq:dar_reduced_obj_single}
\end{equation}
subject to
\begin{equation}
	s_p^{(0)} + Z_0 y \ge 0.
	\label{eq:dar_reduced_nonneg_single}
\end{equation}
In reduced form, \eqref{eq:dar_reduced_obj_single}--\eqref{eq:dar_reduced_nonneg_single}
is a nonnegative weighted least-squares problem, equivalently a convex
quadratic program with linear inequality constraints
\cite{LawsonHanson1995LeastSquares,GoldfarbIdnani1983StrictlyConvexQP}.
The channel reconstruction is then defined by
\begin{equation}
	s^{\ast}
	=
	s_p^{(0)} + Z_0 y^{\ast},
	\label{eq:dar_reconstruction_single}
\end{equation}
where \(y^{\ast}\) solves
\eqref{eq:dar_reduced_obj_single}--\eqref{eq:dar_reduced_nonneg_single}.

Under the single retained \(0\)-order condition, nonnegative feasibility is
automatic whenever \(m_{0}^{\mathrm{fine}}\ge 0\). In this case, the explicit
particular vector \(s_p^{(0)}\) introduced in
\eqref{eq:dar_particular_single} is itself nonnegative, so the exact-retention
affine subspace already contains a nonnegative point. Thus the
single-retention formulation requires no further compatibility condition for
nonnegative feasibility.

The same point may be expressed in terms of the masses
\[
u_i:=p_i s_i,
\qquad
i=1,\dots,N,
\]
for which \(u_i\ge 0\) whenever \(s_i\ge 0\). The retained condition becomes
\[
\sum_{i=1}^{N}u_i = m_{0}^{\mathrm{fine}},
\qquad
u_i\ge 0,
\]
showing that feasibility depends only on the sign of
\(m_{0}^{\mathrm{fine}}\), and not on any additional compatibility relation
with the subgroup total-cross-section nodes.

The reduced objective in \eqref{eq:dar_reduced_obj_single} is also strictly
convex on the affine subspace \eqref{eq:dar_affine_param_single}. As a result,
the admissible reconstruction in the single-retention case is unique. The
reason is that the full matching matrix \(M\) is nonsingular, so the remaining
block \(C\) is nondegenerate once restricted to \(\ker(E_0)\). Equivalently,
the quadratic form induced by \(\widehat C^{\mathsf T}W_s\widehat C\) is
positive definite on \(\ker(E_0)\), and the minimizer of
\eqref{eq:dar_reduced_obj_single}--\eqref{eq:dar_reduced_nonneg_single} is
therefore unique; this is the standard uniqueness mechanism for strictly
convex quadratic programs on a closed convex feasible set
\cite{GoldfarbIdnani1983StrictlyConvexQP,NocedalWright2006NumericalOptimization}.

\paragraph{Two-retention variant.}
A stronger variant retains both the \(0\)-order and \((-1)\)-order fine-state
channel aggregates exactly:
\begin{equation}
	m_{0}^{\mathrm{fine}}
	:=
	\sum_{j=0}^{J}\omega_j r_{x,j},
	\qquad
	m_{-1}^{\mathrm{fine}}
	:=
	\sum_{j=0}^{J}\omega_j r_{x,j}r_{t,j}^{-1}.
	\label{eq:dar_two_moments_fine}
\end{equation}
These two retained quantities are chosen to align with the two endpoint limits emphasized by Chiba, namely the infinite-dilution and zero-dilution endpoints.
Accordingly, define
\begin{equation}
	E_{-1,0}
	:=
	\begin{pmatrix}
		p_1 & \cdots & p_N \\
		p_1\sigma_{t,1}^{-1} & \cdots & p_N\sigma_{t,N}^{-1}
	\end{pmatrix},
	\qquad
	r_{-1,0}
	:=
	\begin{pmatrix}
		m_{0}^{\mathrm{fine}} \\
		m_{-1}^{\mathrm{fine}}
	\end{pmatrix},
	\label{eq:dar_E_two}
\end{equation}
so that the exact-retention constraints become
\begin{equation}
	E_{-1,0}s = r_{-1,0}.
	\label{eq:dar_exact_retention_two}
\end{equation}

Assuming \(\operatorname{rank}(E_{-1,0})=2\), the same null-space reduction as in
the preceding subsection applies. Let \(s_p^{(-1,0)}\) be any particular solution
of \eqref{eq:dar_exact_retention_two}, and let \(Z_{-1,0}\) be a basis matrix for
\(\ker(E_{-1,0})\). Then every exactly retained vector can be written as
\begin{equation}
	s = s_p^{(-1,0)} + Z_{-1,0}y,
	\qquad
	y\in\mathbb{R}^{N-2}.
	\label{eq:dar_affine_param_two}
\end{equation}
As in the single-retention case, this reduction enforces the retained linear
constraints exactly and transfers the optimization to a lower-dimensional
variable.
In the numerical implementation, \(s_p^{(-1,0)}\) is recomputed within each
active-set subproblem after the currently active components have been fixed at
their bounds. This active-set language is standard in both convex quadratic
programming and bounded-variable least-squares methods
\cite{GoldfarbIdnani1983StrictlyConvexQP,StarkParker1995BVLS}.
The reduced problem is therefore
\begin{equation}
	\min_{y\in\mathbb{R}^{N-2}}
	\frac12
	\bigl\|
	W_s^{1/2}\bigl(\widehat C^{(-1,0)}(s_p^{(-1,0)}+Z_{-1,0}y)-\widehat c^{(-1,0)}\bigr)
	\bigr\|_2^2
	\label{eq:dar_reduced_obj_two}
\end{equation}
subject to
\begin{equation}
	s_p^{(-1,0)}+Z_{-1,0}y\ge 0.
	\label{eq:dar_reduced_nonneg_two}
\end{equation}
Although the present implementation uses nonnegativity bounds, the same
reduced formulation extends directly to more general componentwise lower and
upper bounds, as in bounded-variable least-squares formulations
\cite{StarkParker1995BVLS}.

Relative to the single-retention formulation, this two-retention variant
preserves more low-order channel information exactly and, in some energy
groups, may yield smaller local reconstruction errors.
Its trade-off is that nonnegative feasibility is no longer automatic.
To characterize this, define
\begin{equation}
	\mathcal F_{-1,0}^{+}
	:=
	\left\{
	s\in\mathbb R^N:
	\sum_{i=1}^{N} p_i s_i = m_{0}^{\mathrm{fine}},
	\quad
	\sum_{i=1}^{N} p_i \sigma_{t,i}^{-1} s_i = m_{-1}^{\mathrm{fine}},
	\quad
	s_i\ge 0,\ i=1,\dots,N
	\right\}.
	\label{eq:dar_two_constraint_feasible_set}
\end{equation}
Assume \(m_{0}^{\mathrm{fine}}>0\), and define
\[
\lambda_i:=\frac{p_i s_i}{m_{0}^{\mathrm{fine}}},
\qquad
i=1,\dots,N.
\]
Then \(\lambda_i\ge 0\) and \(\sum_{i=1}^{N}\lambda_i=1\), and
\begin{equation}
	\frac{m_{-1}^{\mathrm{fine}}}{m_{0}^{\mathrm{fine}}}
	=
	\sum_{i=1}^{N}\lambda_i \sigma_{t,i}^{-1}.
	\label{eq:dar_ratio_convex_combination}
\end{equation}
Therefore the two-retention formulation admits a nonnegative feasible solution
if and only if
\begin{equation}
	\frac{m_{-1}^{\mathrm{fine}}}{m_{0}^{\mathrm{fine}}}
	\in
	\mathrm{conv}\!\left\{
	\sigma_{t,1}^{-1},\dots,\sigma_{t,N}^{-1}
	\right\},
	\label{eq:dar_conv_condition_two}
\end{equation}
or equivalently,
\begin{equation}
	\min_{1\le i\le N}\sigma_{t,i}^{-1}
	\le
	\frac{m_{-1}^{\mathrm{fine}}}{m_{0}^{\mathrm{fine}}}
	\le
	\max_{1\le i\le N}\sigma_{t,i}^{-1}.
	\label{eq:dar_interval_condition_two}
\end{equation}

Accordingly, the single-retention formulation is adopted as the default because
its nonnegative feasible set is always nonempty whenever
\(m_{0}^{\mathrm{fine}}\ge 0\), while the two-retention formulation is kept as a
stronger alternative.
The latter imposes a lower-dimensional exact-retention affine subspace and may,
in some groups, yield smaller residuals or smaller local errors, but only when
the retained ratio is compatible with the compressed subgroup total nodes.

\paragraph{Overall computational procedure.}
For each energy group, the overall construction proceeds as follows.
\begin{enumerate}
    \item Form the fine-state data
          \(\{(r_{t,j},r_{x,j},\omega_j)\}_{j=0}^{J}\).

    \item Construct the transformed total-cross-section realization
          \(\mu_M=\sum_{j=0}^{J} w_j\delta_{z_j}\).

    \item Apply \(N\) steps of the symmetric Lanczos process to
          \((A,v_0)\) and assemble the projected Jacobi matrix \(J_N\).

    \item Compute the eigendecomposition
          \(J_N=Q\Lambda Q^{\mathsf T}\), and extract
          \[
          \sigma_{t,i}=\hat z_i^{\,1/b},\qquad
          p_i=Q_{0i}^2,\qquad i=1,\dots,N.
          \]

    \item Evaluate the reaction-channel orthogonal-basis coefficients
          \(b_k\), \(k=0,\dots,N-1\).

    \item Form the retained data \((E,r)\) and the remaining matching system.
          In the default formulation, \((E,r)=(E_0,r_0)\); in the
          two-retention variant, \((E,r)=(E_{-1,0},r_{-1,0})\).

    \item Construct a particular vector \(s_p\) such that \(Es_p=r\).

    \item Compute a basis matrix \(Z\) for \(\ker(E)\), and write
          \[
          s=s_p+Zy.
          \]

\item Solve the reduced nonnegative weighted least-squares problem for \(y\)
      on the exact-retention affine subspace.

\item Recover \(s=s_p+Zy\), and combine
      \(\{(\sigma_{t,i},p_i,\sigma_{x,i})\}_{i=1}^{N}\)
      in the subgroup folding formula.
\end{enumerate}

\section{Numerical Results}

We now examine the admissible reconstructions for the representative
\({}^{238}\mathrm{U}\) capture case based on pointwise cross sections
reconstructed from ENDF/B-VIII.1. Attention is restricted to those energy
groups in which the full-matching solution violates componentwise
nonnegativity. Table~\ref{tab:violating_groups_one_two_constraints} summarizes
the main aggregate comparison through the \(0.95\)-quantile absolute relative
error in the effective cross section and the Euclidean distance to the original
full-matching solution. The figures that follow provide complementary
diagnostics, including cumulative profiles, mixed-moment errors, and
dilution-dependent response errors.

\begin{table}[htbp]
\centering
\caption{Violating energy groups and corresponding \(0.95\)-quantile absolute relative errors in the effective cross section for the full-matching, single-retention, and double-retention reconstructions.}
\label{tab:violating_groups_one_two_constraints}
\scriptsize
\setlength{\tabcolsep}{4pt}
\begin{tabular}{ccccccc}
\hline
\(N\) & Group & \(\varepsilon_{\mathrm{full}}^{0.95}\)
& \(\varepsilon_{\mathrm{single}}^{0.95}\)
& \(\|s_{\mathrm{single}}-s_{\mathrm{full}}\|_2\)
& \(\varepsilon_{\mathrm{double}}^{0.95}\)
& \(\|s_{\mathrm{double}}-s_{\mathrm{full}}\|_2\) \\
\hline
5  & 2  & \(9.98\times 10^{-4}\)  & \(9.88\times 10^{-4}\)  & \(3.95\times 10^{-2}\) & \(9.87\times 10^{-4}\)  & \(3.96\times 10^{-2}\) \\
5  & 19 & \(1.75\times 10^{-4}\)  & \(6.44\times 10^{-4}\)  & \(2.49\times 10^{0}\)  & \(6.53\times 10^{-4}\)  & \(2.52\times 10^{0}\) \\
10 & 18 & \(3.45\times 10^{-7}\)  & \(8.90\times 10^{-7}\)  & \(1.79\times 10^{0}\)  & \(8.90\times 10^{-7}\)  & \(1.79\times 10^{0}\) \\
10 & 22 & \(4.54\times 10^{-13}\) & \(5.69\times 10^{-13}\) & \(1.15\times 10^{-1}\) & \(6.55\times 10^{-13}\) & \(3.06\times 10^{-1}\) \\
20 & 22 & \(4.54\times 10^{-13}\) & \(6.77\times 10^{-13}\) & \(4.27\times 10^{-1}\) & \(2.40\times 10^{-6}\)  & \(2.55\times 10^{0}\) \\
30 & 22 & \(4.53\times 10^{-13}\) & \(2.03\times 10^{-12}\) & \(6.93\times 10^{-1}\) & \(1.96\times 10^{-11}\) & \(2.14\times 10^{0}\) \\
30 & 32 & \(1.10\times 10^{-13}\) & \(8.25\times 10^{-7}\)  & \(6.26\times 10^{0}\)  & \(1.82\times 10^{-7}\)  & \(2.87\times 10^{0}\) \\
50 & 18 & \(4.94\times 10^{-14}\) & \(6.48\times 10^{-8}\)  & \(8.06\times 10^{0}\)  & \(2.46\times 10^{-8}\)  & \(9.66\times 10^{0}\) \\
50 & 22 & \(4.54\times 10^{-13}\) & \(2.03\times 10^{-12}\) & \(9.30\times 10^{-1}\) & \(1.27\times 10^{-9}\)  & \(4.38\times 10^{0}\) \\
50 & 25 & \(4.29\times 10^{-13}\) & \(6.44\times 10^{-9}\)  & \(4.74\times 10^{1}\)  & \(1.28\times 10^{-10}\) & \(2.39\times 10^{1}\) \\
\hline
\end{tabular}
\end{table}

\begin{table}[htbp]
\centering
\caption{Upper and lower energy bounds of the violating resonance groups listed in Table~\ref{tab:violating_groups_one_two_constraints}.}
\label{tab:cropped_es2_selected_group_bounds}
\begin{tabular}{ccc}
\toprule
Group & Upper boundary (eV) & Lower boundary (eV) \\
\midrule
2  & $7.465850\times10^{3}$ & $6.112520\times10^{3}$ \\
18 & $8.322180\times10^{2}$ & $7.485170\times10^{2}$ \\
19 & $7.485170\times10^{2}$ & $6.772870\times10^{2}$ \\
22 & $6.128340\times10^{2}$ & $6.000990\times10^{2}$ \\
25 & $5.771460\times10^{2}$ & $5.392040\times10^{2}$ \\
32 & $3.535750\times10^{2}$ & $3.353230\times10^{2}$ \\
\bottomrule
\end{tabular}
\end{table}

Two general observations follow from
Table~\ref{tab:violating_groups_one_two_constraints}. First, the violating
groups constitute only a limited subset of the tested cases, so the admissible
reconstruction acts only locally. Second, even on these groups, the
full-matching solution is often already highly accurate at the response level,
with \(\varepsilon_{\mathrm{full}}^{0.95}\) frequently lying between
\(10^{-13}\) and \(10^{-7}\).

The comparison between the single-retention and double-retention
reconstructions shows a trade-off rather than a uniform ordering. In some
violating groups, the double-retention variant gives a slightly smaller
\(0.95\)-quantile response error. However, this advantage is not systematic:
in several other groups, it yields substantially larger response errors and a
larger departure from the full-matching solution in the \(\ell_2\) norm.
Overall, the single-retention reconstruction is associated with the more
moderate perturbation of the full-matching solution across the tested cases.

This contrast is particularly evident for the recurrent violating group 22. As
\(N\) increases from \(10\) to \(20\), \(30\), and \(50\), the
single-retention reconstruction shows a clear increase in
\(\|s_{\mathrm{single}}-s_{\mathrm{full}}\|_2\), with a milder but visible
increase in the corresponding \(0.95\)-quantile response error. The
double-retention reconstruction exhibits an even less regular pattern on the
same group, including much larger error excursions in some higher-order cases.

\begin{figure}[htbp]
\centering
\includegraphics[width=0.78\textwidth]{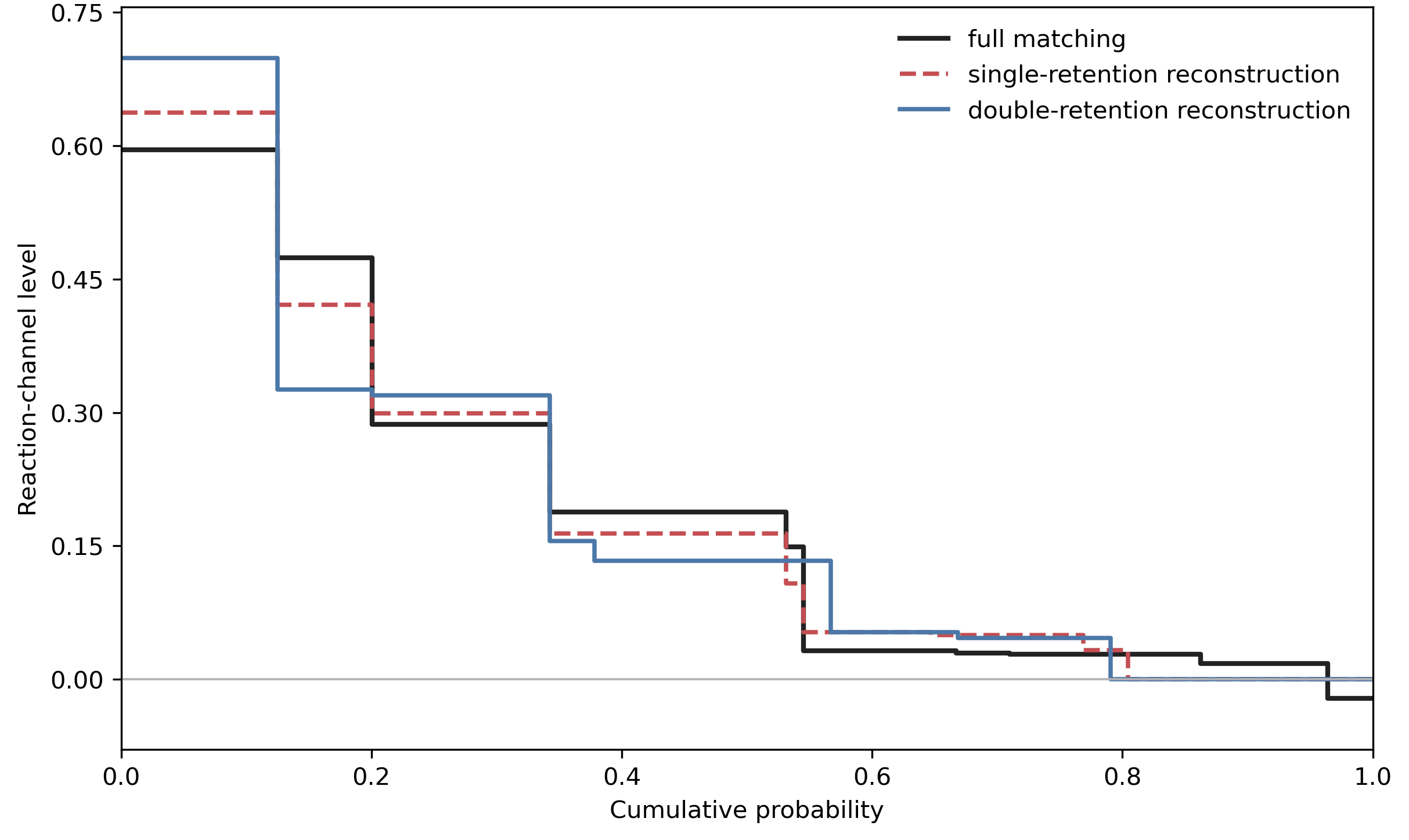}
\caption{Cumulative-probability representation of the reaction-channel subgroup levels in energy group 22 for \(N=10\). The black curve is the original full-matching reconstruction, while the red dashed and blue solid curves denote the single-retention and double-retention reconstructions, respectively. Both admissible reconstructions remove the nonphysical negative tail of the full-matching solution on the same compressed cumulative-probability partition, while exhibiting different degrees of redistribution in the reaction-channel levels.}
\label{fig:N10_group22_cumprob_compare}
\end{figure}

\begin{figure}[htbp]
\centering
\includegraphics[width=0.78\textwidth]{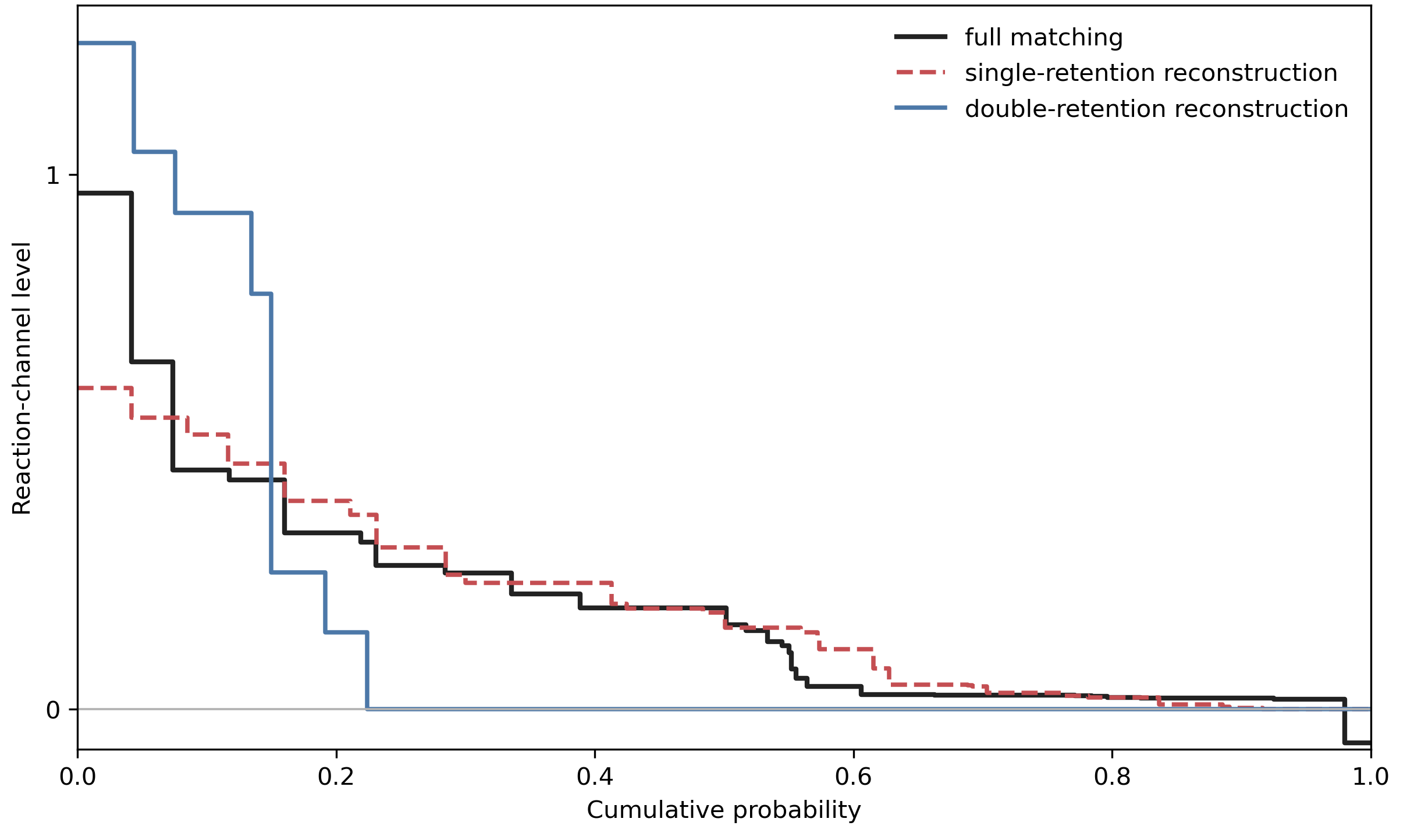}
\caption{Reaction-channel levels in cumulative-probability form for energy group 22 at \(N=30\). The full-matching reconstruction develops a negative tail, whereas the single-retention and double-retention reconstructions restore nonnegativity on the same compressed probability partition.}
\label{fig:group22_cumprob_compare}
\end{figure}

The cumulative-probability plots in
Figures~\ref{fig:N10_group22_cumprob_compare} and
\ref{fig:group22_cumprob_compare} illustrate the geometric effect of the
admissible reconstruction. In both representative cases, the full-matching
reconstruction develops a negative tail, whereas both admissible
reconstructions eliminate that tail on the same compressed probability
partition. At \(N=10\), the single-retention and double-retention
reconstructions remain relatively close to one another, indicating that
admissibility restoration is achieved through a modest redistribution of the
channel levels. At \(N=30\), the difference between the two admissible
reconstructions becomes much more pronounced: both restore nonnegativity, but
the double-retention variant produces a visibly stronger redistribution of the
cumulative profile. This is consistent with the larger \(\ell_2\) distances
reported in Table~\ref{tab:violating_groups_one_two_constraints}.

\begin{figure}[htbp]
\centering
\includegraphics[width=0.78\textwidth]{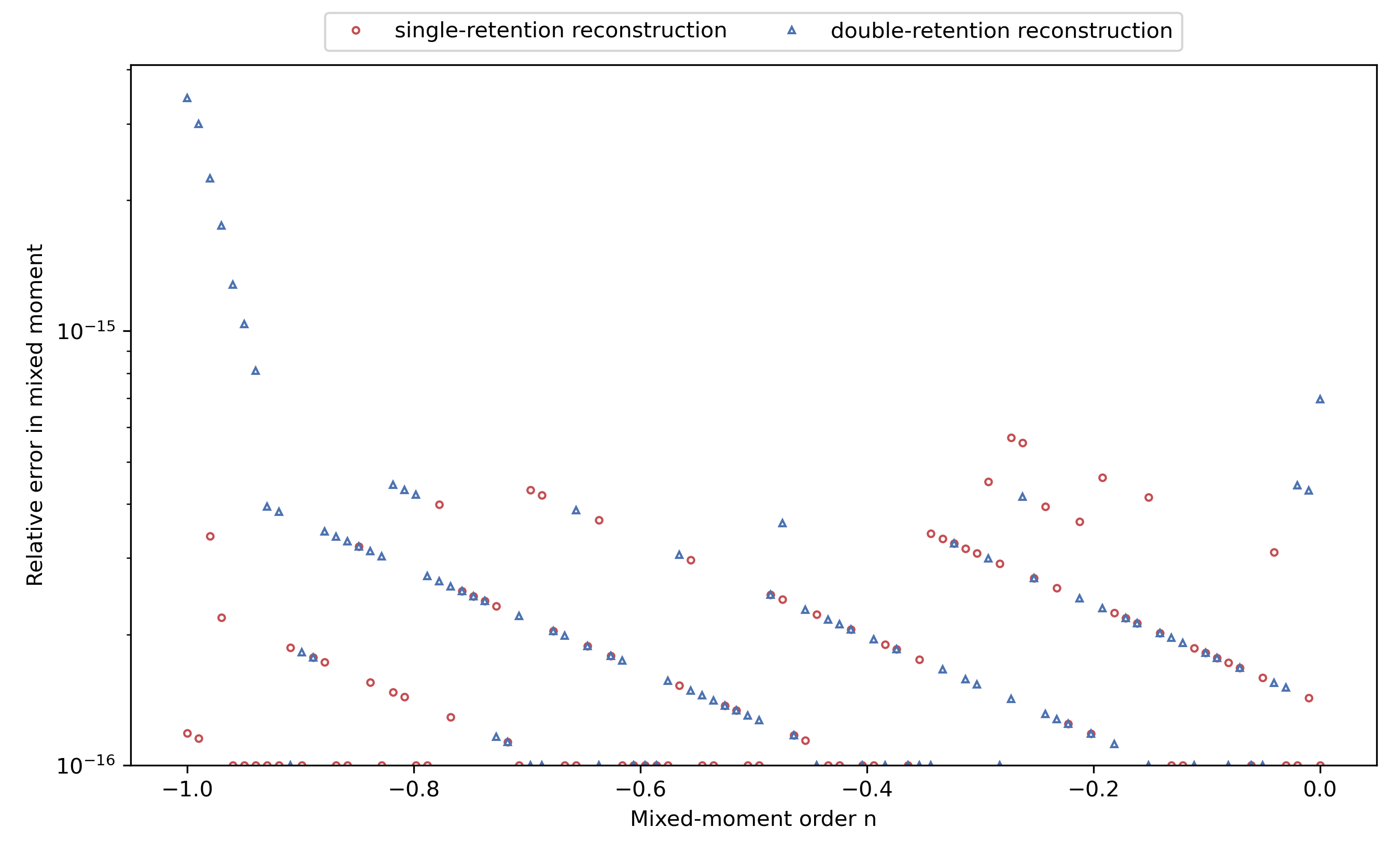}
\caption{Relative errors in the mixed moments as functions of the sampled mixed-moment order \(n\in[-1,0]\) for energy group 18 at \(N=10\). The single-retention and double-retention reconstructions both exhibit mixed-moment errors near machine precision over the sampled interval. For logarithmic display, zero values are shown at the plotting floor \(10^{-16}\).}
\label{fig:moment_errors_compare_scatter}
\end{figure}

\begin{figure}[htbp]
\centering
\includegraphics[width=0.78\textwidth]{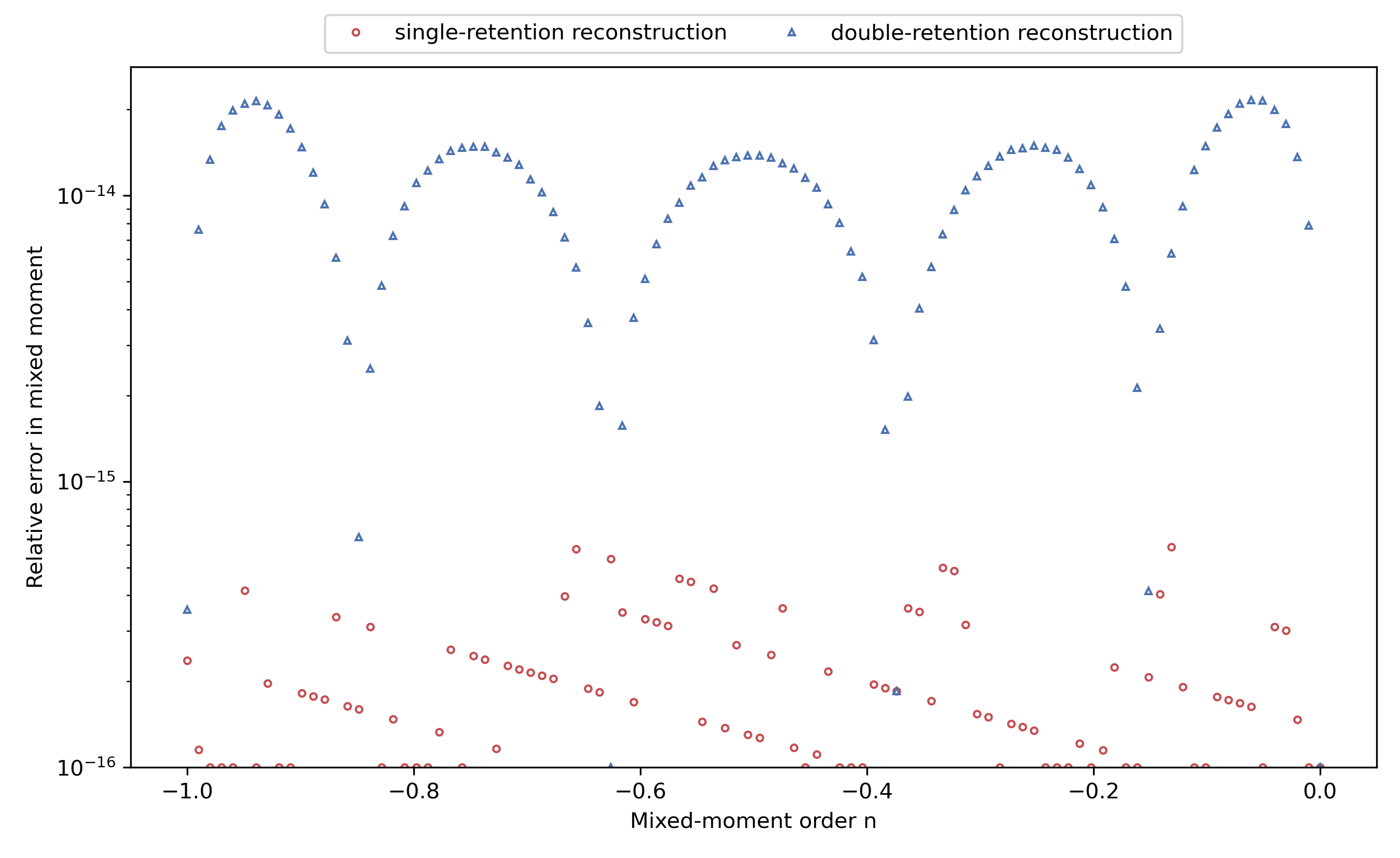}
\caption{Relative mixed-moment errors over the sampled order interval \(n\in[-1,0]\) for energy group 22 at \(N=30\). The single-retention and double-retention reconstructions exhibit distinct retained-moment behavior on this group: the single-retention reconstruction remains closer to the plotting floor across the sampled orders, whereas the double-retention reconstruction shows larger moment errors, though still at a very small absolute level. For logarithmic display, zero values are shown at the plotting floor \(10^{-16}\).}
\label{fig:moment_errors_compare_scatter_g22_n30}
\end{figure}

The mixed-moment diagnostics in
Figures~\ref{fig:moment_errors_compare_scatter} and
\ref{fig:moment_errors_compare_scatter_g22_n30} provide a complementary view.
For group 18 at \(N=10\), both admissible reconstructions maintain relative
mixed-moment errors near machine precision throughout the sampled interval
\(n\in[-1,0]\). For group 22 at \(N=30\), however, the distinction between the
two variants becomes clearer: the single-retention reconstruction remains
closer to the plotting floor over most of the sampled orders, whereas the
double-retention reconstruction exhibits larger mixed-moment errors, although
still at a very small absolute level. Thus, the stronger exact retention
imposed by the double-retention variant does not translate into uniformly
better behavior over the sampled mixed-moment interval.

\begin{figure}[htbp]
\centering
\includegraphics[width=0.78\textwidth]{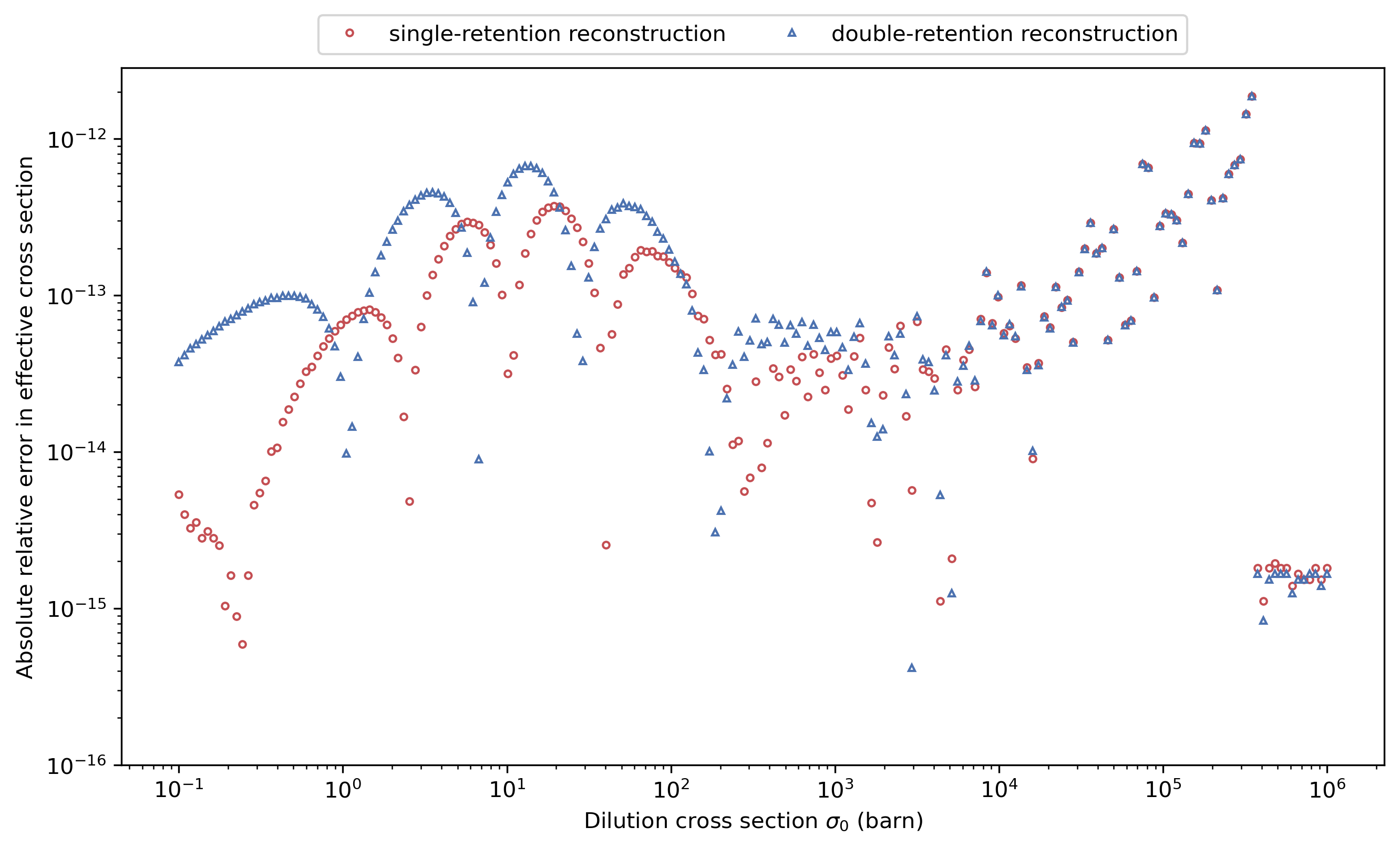}
\caption{Absolute relative errors in the effective cross section as functions of the dilution cross section \(\sigma_0\) for energy group 22 at \(N=10\). The single-retention and double-retention reconstructions exhibit similar overall error profiles over the sampled dilution range, with visible local differences in the intermediate-\(\sigma_0\) region.}
\label{fig:N10_sigma0_error_compare_group22_absrel}
\end{figure}

\begin{figure}[htbp]
\centering
\includegraphics[width=0.78\textwidth]{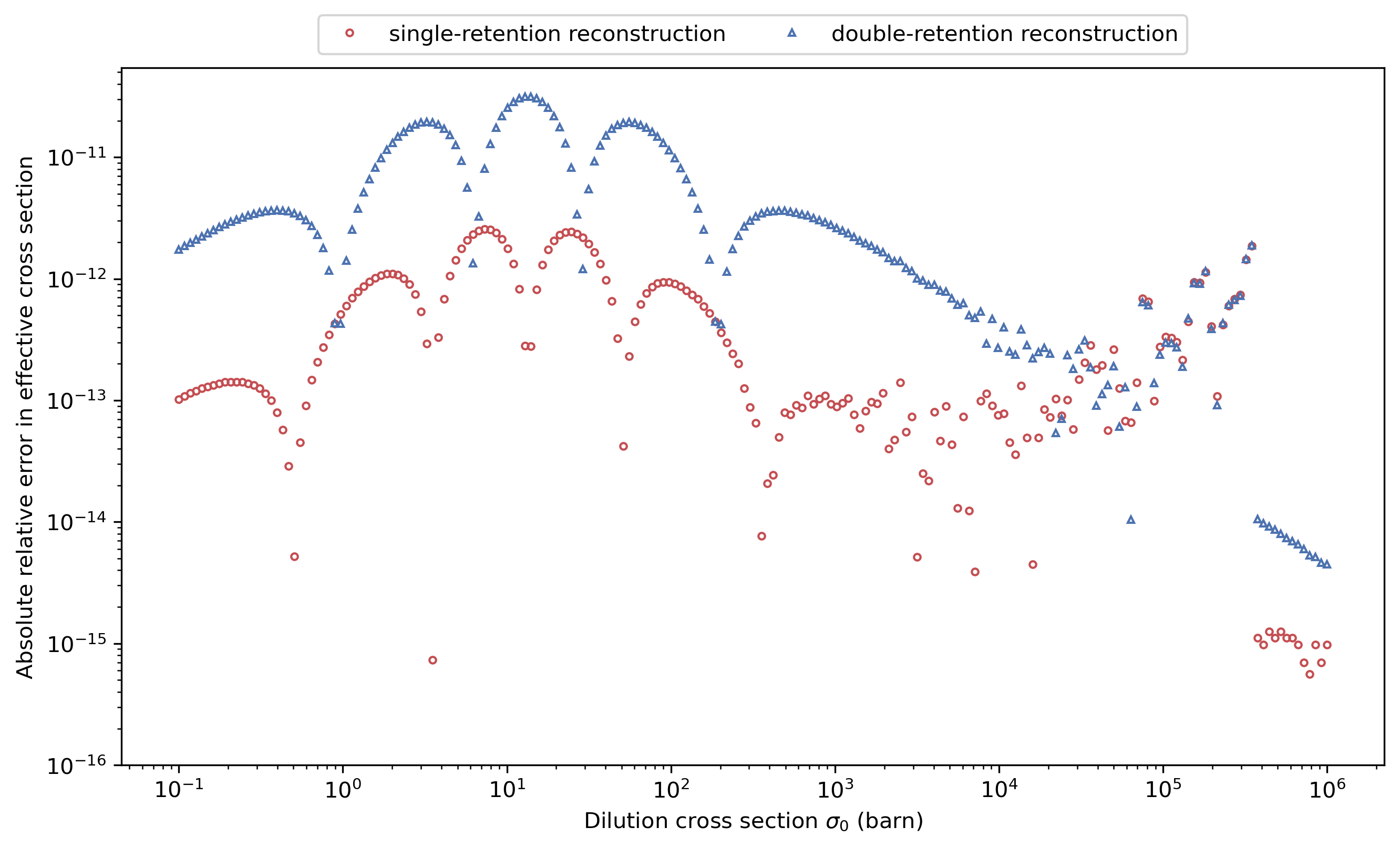}
\caption{Absolute relative errors in the effective cross section as functions of the dilution cross section \(\sigma_0\) for energy group 22 at \(N=30\). The single-retention and double-retention reconstructions show markedly different error levels over the sampled dilution range, with the single-retention reconstruction remaining lower over much of the interval.}
\label{fig:N30_sigma0_error_compare_group22_absrel}
\end{figure}

A similar pattern is observed in the dilution-dependent effective-cross-section
errors shown in
Figures~\ref{fig:N10_sigma0_error_compare_group22_absrel} and
\ref{fig:N30_sigma0_error_compare_group22_absrel}. For group 22 at \(N=10\),
the single-retention and double-retention reconstructions produce broadly
similar error profiles, with only localized differences in the intermediate
\(\sigma_0\) range. At \(N=30\), by contrast, the separation becomes much more
distinct: the single-retention reconstruction remains below the
double-retention reconstruction over a substantial portion of the sampled
dilution interval. Together with the \(\ell_2\) distances reported in
Table~\ref{tab:violating_groups_one_two_constraints}, this indicates that, on
this representative difficult group, the stronger two-retention variant does
not lead to a more favorable overall response profile once admissibility is
imposed.

\section{Conclusion}

We have considered the reconstruction of reaction-channel subgroup levels on fixed total-subgroup nodes and probabilities. Although the full-matching reconstruction is uniquely determined, it does not in general preserve componentwise nonnegativity. We therefore formulated an admissible reconstruction that enforces nonnegativity while retaining selected low-order channel information exactly.

For the single-retention formulation, nonnegative feasibility is automatic whenever the retained \(0\)-order aggregate is nonnegative. The two-retention variant preserves additional low-order information, but its feasibility requires an additional compatibility condition with the fixed total-subgroup nodes.

Numerical results for the representative \(^{238}\mathrm{U}\) capture case based on ENDF/B-VIII.1 show that nonnegativity violations are confined to a limited subset of energy groups. On these groups, the admissible reconstruction restores nonnegativity, with some deterioration in response accuracy relative to full matching. Across the present tests, the single-retention reconstruction shows the more stable overall behavior, while the two-retention variant remains useful as a stronger alternative when its additional retained information is desired and the corresponding feasibility condition is satisfied.
	
	\bibliographystyle{unsrt}
	\bibliography{ref_SPR}
	
\end{document}